\documentclass[a4paper,12pt]{article}

\usepackage{graphicx}
\usepackage{amsfonts}
\usepackage{amsmath}

\renewcommand{\phi}{\varphi}

\begin{document}
\def\bibname{References}
\def\proof{\textbf{Proof. }}
\newtheorem{pro*}{Proposition}
\renewcommand{\thepro*}{}
\thispagestyle{empty}
\begin{center}
{\Large {\bf Interaction between Kirchhoff vortices and point vortices in an ideal
fluid}}\\
\bigskip\bigskip
{\bf Alexey V. Borisov, Ivan S. Mamaev}\\
Institute of Computer Science\\
Udmurt State University, Izhevsk, Russia\\
E-mail: borisov@ics.org.ru\\
\end{center}
\renewcommand{\theequation}{\arabic{equation}}
\renewcommand{\thesubsection}{\arabic{subsection}}

{\bf Abstract.} We consider the interaction of two vortex patches (elliptic Kirchhoff
vortices) which move in an unbounded volume of an ideal incompressible
fluid. A moment second-order model is used to describe the interaction.
The case of integrability of a Kirchhoff vortex and a point vortex by the
variable separation method is qualitatively analyzed. A new case of
integrability of two Kirchhoff vortices is found. A reduced form of
equations for two Kirchhoff vortices is proposed and used to analyze their
regular and chaotic behavior.

\newpage

\section{Introduction.}\mbox{}

The simplest example of planar vortex motion of an ideal fluid, other than
that described by the point vortex model, was suggested by Kirchhoff \cite{cccbib2}.
He established that an elliptical vortex patch with
semiaxes of $a$, $b$ and a uniform vorticity $\omega$ inside uniformly
rotates around its center with an angular velocity of
$\Omega=(\omega\lambda)/(1+\lambda)^2$, $\lambda=a/b$. In this
rotation, the fluid particles are involved in the absolute motion with a
double angular velocity (Lamb, 1932). Lord Kelvin (1880) and Love (1893)
showed that the Kirchhoff vortex is neutrally stable if and only if $a/b<3$.

S.\,A.\,Chaplygin in 1899 generalized Kirchhoff's solution by introducing
a uniform background vorticity into the unbounded fluid surrounding the
elliptic vortex (this is the so-called Couette simple shearing motion). He
established that the vortex will rotate with a certain angular velocity
and change its contour (pulsate) in accordance with a certain law, which
he obtained by integrating a system of two nonlinear differential
equations. He also analyzed in detail the behavior of pressure in the
fluid as a function of time.

Kida \cite{cccbib13} and Neu \cite{cccbib14} generalized
Chaplygin's solution while being unaware of his work \cite{cccbib6}.
Chaplygin's works in many fields of mechanics are poorly known outside the
Russian speaking world. Forgotten Chaplygin's works on two-dimensional
vortex structures are discussed in the recent historical review by
Meleshko and van Heijst \cite{cccbib23}. A superposition of the solutions obtained by Kida and Neu
is presented in the book by Newton\cite{cccbib16}. It was found that the dynamics
of a Kirchhoff vortex can be reduced to a one-degree-of-freedom
Hamiltonian system  if the velocity of the external flow can be described
by
\begin{equation}\label{ccceq1}
\begin{gathered}
{\bf u}={\bf u}_1+{\bf u}_2+{\bf u}_3,\\
{\bf u}_1=(\gamma_1x,-\gamma_2y,0),\\
{\bf u}_2=(0,0,\gamma_3z),\\
{\bf u}_3=(-\gamma_4y,\gamma_4x,0)
\end{gathered}
\end{equation}
under the condition $\gamma_1-\gamma_2+\gamma_3=0$ derived from the
incompressibility condition, and. It is assumed that the elliptic patch is
the section of the elliptical cylinder determined by the plane $xy$.
Equation \eqref{ccceq1} yields Chaplygin's solution when
${\gamma_1=\gamma_2=\gamma_4\ne0}$, $\gamma_3=0$; Kida's solution when
$\gamma_3=0$, $\gamma_1=\gamma_2$; and Neu's solution when $\gamma_4=0$.
Note that $\bf {u}_1$ corresponds to the uniform deformation field (which
is induced at a distance by one point vortex \cite{cccbib8}, \cite{cccbib7}); the
field $\bf {u}_2$ corresponds to the extension along $z$ axis, and the
field $\bf {u}_3$ corresponds to uniform background rotation. The
Hamiltonian system that describes this case has the form
\begin{equation}\label{ccceq2}
\dot\lambda=-\frac{\partial
H}{\partial\theta},\quad\dot\theta=\frac{\partial
H}{\partial\lambda},\quad
H(\lambda,\theta)=\ln\frac{(1+\lambda)^2}\lambda-\frac12\frac{\gamma_1+\gamma_2}{\omega(t)}
\Bigl(\lambda-\frac1\lambda\Bigr)\sin2\theta,
\end{equation}
where the dot denotes differentiation with respect to the new time
$\tau=\omega(t)(\lambda^2/(\lambda^2-1))t$, while
$\omega(t)=\omega_0e^{(\gamma_2-\gamma_1)t}$, and $\omega_0$ is the
initial vorticity in the Kirchhoff vortex. If $\gamma_1=\gamma_2$, the
system \eqref{ccceq2} is autonomous and can be explicitly integrated.
Qualitative analysis of this system is given in \cite{cccbib13}, \cite{cccbib14},
\cite{cccbib8}, \cite{cccbib7}.

When the coefficients in \eqref{ccceq1} are functions of time (for
example, periodical), we obtain (at $\gamma_1=\gamma_2$) a Hamiltonian
system with one and a half degrees of freedom with periodic perturbations,
which was studied  from the viewpoint of
splitting of separatrices and the appearance of stochastic behavior
{\cite{cccbib9}, \cite{cccbib15}. The
problem of advection of a passive particle of the fluid in the Kida flow
was discussed in \cite{cccbib22}. Examination of Poincar\'{e} sections
allows us to conclude that the advection is chaotic, i.\,e., Lagrangian
turbulence exists. The stability of the elliptic vortex in a Kida flow
(uniform deformation) was studied in Ref. \cite{cccbib21}.

Finally, we will mention the generalized Kirchhoff solution obtained by Polvani and
Flierl \cite{cccbib24}  and corresponding to a family of embedded confocal
elliptic vortices with the appropriate distribution of vorticity. In that
work, the stability conditions, which generalize the Kelvin--Love
conditions, were obtained for a system of two confocal elliptic vortices.

\section{Moment model of Kirchhoff vortex interaction (vortex patch dynamics).}

The second-order moment model \cite{cccbib4} represents a higher level of
approximation in the description of hydrodynamic vorticity as compared
with the point vortex model. This model considers elliptic patches with a
specified vorticity, which move in a two-dimensional incompressible fluid.
This theory was suggested by Melander, Zabusky, and Stychek (MZS model) in
Ref. \cite{cccbib4}. Rotation of two Kirchhoff vortices in the presence of
central symmetry is analyzed in \cite{cccbib11}. The nonsymmetrical
situation of interaction between two vortex patches is discussed in
\cite{cccbib10}, where the same moment model, which describes merging of
vortices, is compared with a pseudospectral model (in which Euler
equations with a weak dissipation are solved).

Let us describe this model in more detail. The second-order moment model,
describing Kirchhoff vortex interactions, is derived from two basic
assumptions:

\begin{enumerate}

\item the distance between vortices in the process of evolution is much greater than the vortex size, therefore the vortices will maintain their elliptic shape;

\item the third and higher moments in the expansion of the Hamiltonian can be neglected.

\end{enumerate}


Under these assumptions, the equations of motion of elliptic vortices can
be written in the Hamiltonian form with nonlinear Poisson brackets
\cite{cccbib4}
\begin{equation}
\label{in1}
\begin{gathered}
\dot x_k=\{x_k,H\},\quad \dot y_k=\{y_k,H\},\quad
\dot\varphi_k=\{\varphi_k,H\},\quad \dot\lambda_k=\{\lambda_k,H\},\\
\{x_i,y_j\}=\frac{1}{\Gamma_i}\delta_{ij},\quad
\{\varphi_i,\lambda_j\}=\frac{8\pi}{\Gamma_iS_i}\frac{\lambda_i^2}{1-\lambda_i^2}\delta_{ij},
\end{gathered}
\end{equation}
and the Hamiltonian is
\begin{equation}
\label{in3}
\begin{gathered}
H=H_1+H_2+H_3,\\
H_1=-\frac1{8\pi}\sum\limits_{k=1}^N\Gamma_k^2\ln\frac{(1+\lambda_k)^2}{4\lambda_k},\quad
H_2=-\frac1{8\pi}\mathop{{\sum}'}\limits_{k,p}^N\Gamma_k\Gamma_p\ln M_{kp},\\
H_3=-\frac1{32\pi^2}\mathop{{\sum}'}\limits_{k,p}^N\frac{\Gamma_k\Gamma_p}{M_{kp}}
\left(S_p\frac{1-\lambda_p^2}{\lambda_p}\cos(2(\theta_{kp}-\varphi_p)){}\right.\\
\left.{}+S_k\frac{1-\lambda_k^2}{\lambda_k}\cos(2(\theta_{kp}-\varphi_k))\right),
\end{gathered}
\end{equation}
where $\Gamma_k$, $S_k$ are the total intensity and the area of the
elliptic vortex with index~$k$; $M_{kp}$ is the squared distance between
the centers of the $k$-th and $p$-th vortices $\left(M_{kp}=(x_k-x
_p)^2+(y_k-y_p)^2\right)$; $\varphi_k$ is the slope of the $k$-th ellipse
with respect to the $x$ axis; and $\theta_{kp}$ is the angle between the
$x$-axis and the straight line connecting the centers of the $k$-th and
$p$-th ellipses (see Fig.~1).

The area of each ellipse remains constant ($S_k={\rm const}$)  by virtue
of Kelvin's theorem on the conservation of circulation in an ideal
medium \cite{97}, \cite{Lamb}; therefore, these areas are parameters of the model
considered.

The components of the Hamiltonian have the following physical meaning:

$H_1$ describes the action of an elliptic vortex on itself, see
\eqref{ccceq2};

$H_2$ describes the interaction between equivalent point vortices;

$H_3$ describes the interaction between vortices with second-order moments
taken into account.

In addition to the Hamiltonian $H$, equations (\ref{in1}) have a {\it
noncommutative} set of first integrals
\begin{equation}
\label{in4} Q=\sum\limits_k^N\Gamma_kx_k,\;P=\sum\limits_k^N\Gamma_ky_k,\;
I=\sum\limits_k^N\Gamma_k\left[x_k^2+y_k^2+\frac{S_k}{4\pi}\frac{1+\lambda_k^2}{\lambda_k}
\right],
\end{equation}
which correspond to the translational and rotational invariance of the
system in the absolute space.

Integrals $Q$, $P$, $I$ satisfy the following commutation relations:
\begin{equation}
\label{ccceq61_2} \{Q,P\}=\sum\limits_{i=1}^N\gamma_i,\quad
\{P,I\}=-2Q,\quad \{Q,I\}=2P.
\end{equation}
Therefore these relations are not sufficient even for the integrability of
a system of two Kirchhoff vortices; this system will be further reduced to
a system with two degrees of freedom. However, the problem of the dynamics
of a Kirchhoff vortex and a point vortex (a system with three degrees of
freedom) is integrable.

A more general model of interaction between elliptic vortex patches was
suggested by D.\,Dritschel and B.\,Legras \cite{cccleg}, \cite{cccdri}. In some
sense, this is an intermediate model between the MZS-model and an exact
description determined by the contour dynamics method (in the contour
dynamics method, the elliptic shape of the patch does not persist, while
the model \cite{cccleg}, \cite{cccdri} is derived with no allowance made for the
velocity field responsible for the nonelliptic part of the interaction).
However, this model is more complex and in many cases it is sufficient to
use the MZS-model.

\section{Interaction of a Kirchhoff vortex with $N$ point vortices. Integrable case for $N=1$.}\label{alg}

Let us denote the coordinates of the center of the elliptic vortex by
$(x_0,y_0)$ and those of the point vortices, by $(x_i,y_i)$,
$i=1,\ldots,N$; the equations describing the dynamics of this system also
can be written in Hamiltonian form with a Poisson bracket and the
Hamiltonian in the form
\begin{equation}
\label{eq1}
\begin{gathered}
\{x_i,y_i\}=\Gamma_i^{-1}\delta_{ij},\quad i,j=0,1,\ldots, N,\quad\{\varphi,\lambda\}
=\frac{8\pi}{\Gamma_0S}\frac{\lambda^2}{1-\lambda^2},\\
H=H_1+H_2+H_3,\\
H_1=-\frac1{8\pi}\Gamma_0^2\ln{\frac{(1+\lambda)^2}{4\lambda}},\quad
H_2=-\frac{1}{8\pi}\mathop{{\sum}'}_{k,p=0}^N\Gamma_k\Gamma_p\ln M_{kp},\\
H_3=-\frac{\Gamma_0S}{16\pi^2}\sum_{k=1}^N\frac{\Gamma_k}{M_{k0}}
\frac{1-\lambda^2}{\lambda}\cos2(\theta_k-\varphi),
\end{gathered}
\end{equation}
where $\Gamma_0$~is the intensity of the Kirchhoff vortex with the
semiaxes ratio of $\lambda$ with the angle~$\varphi$, determining its
orientation (see Fig.~1), $S$ is the area of the ellipse,
$\theta_k$~is the angle between the $x$-axis and the straight line
connecting the center of the Kirchhoff vortex and the $k$-th point vortex,
and $M_{kp}=(x_k-x_p)^2+(y_k-y_p)^2$. Hereafter, we will assume without
loss of generality that the intensity of the Kirchhoff vortex is positive
($\Gamma_0>0$).

The integrals of motion corresponding to the group $E(2)$ of motions of
the plane are determined by relationships \eqref{in4}, where it should be
assumed that $S_0=S$, and $S_i=0$, $i=1,\ldots, N$; their commutation
relationships are analogous to \eqref{ccceq61_2}. As a consequence of the
existence of integrals, we obtain \cite{cccbib12}:

\begin{pro*}
The system of an interacting Kirchhoff vortex and one point vortex $(N=1)$
is completely integrable.
\end{pro*}

This was first shown by Lebedev \cite{cccbib12} and somewhat later was established
independently by Riccardi and Piva \cite{cccbib5}. Here we will give a geometric analysis of
the motion of vortices to improve the results of \cite{cccbib12}.

For explicit integration and qualitative analysis, we perform a reduction
to one degree of freedom. Let us consider new (relative) variables
\begin{equation}
\label{eqq2} \psi=2(\theta-\varphi),\quad
\rho=\frac12c\Bigl(\lambda+\frac1\lambda\Bigr),\quad
z=M_{10}=(x_1-x_0)^2+(y_1-y_0)^2,
\end{equation}
where $c=\Gamma_0S/8\pi$. All these functions~\eqref{eqq2} commute
with the integrals~\eqref{in4}, i.\,e., they are invariants of the group
of motions of the plane $E(2)$ and are closed with respect to the Poisson
bracket
\begin{equation}
\label{eqq3} \{\psi,\rho\}=1,\quad
\{\psi,z\}=-4(\Gamma_0^{-1}+\Gamma_1^{-1}),\quad \{\rho,z\}=0.
\end{equation}
The Poisson structure~\eqref{eqq3} has a linear Kazimir function
\begin{equation}
\label{eqq4} D=\Gamma_0z+4(1+\alpha)\rho, \quad
\alpha=\Gamma_0^{-1}\Gamma_1.
\end{equation}

Eliminating $z$ with the use of equation \eqref{eqq4}, we obtain the
Hamiltonian of the system (up to the constant) with one degree of freedom:
\begin{equation}
\label{eq5}
H_{\mp}=\frac{\Gamma_0^2}{8\pi}\left(\!-\!\ln(c+\rho){}-2\alpha\ln
(D-4(1+\alpha)\rho)\!\mp8\alpha\frac{\sqrt{\rho^2-c^2}}{D{}-4(1+\alpha)\rho}\cos\psi\right),
\end{equation}
where, according to~\eqref{eqq3}, the variables $\psi$, $\rho$ are
canonical.

The different signs in the Hamiltonian~\eqref{eq5} appear due to the
nonuniqueness of the inverse transformation~\eqref{eqq2} for
$\rho(\lambda)$, which has the form
\begin{equation}
\label{eq6} \lambda=\left\{
\begin{gathered}
\frac{\rho-\sqrt{\rho^2-c^2}}{c}, \quad 0<\lambda\le1\\
\frac{\rho+\sqrt{\rho^2-c^2}}{c}, \quad \lambda>1
\end{gathered} \right.
\end{equation}
(the upper sign in~\eqref{eq5} corresponds to the case $\lambda<1$).

The reduction by (to two degrees of freedom) for arbitrary $N$ can be made
in a similar manner. The new variables for the reduced system can be
chosen as follows:
$$
\begin{gathered}
\psi_k=2(\theta_k-\varphi),\quad \rho=\frac12c(\lambda+\lambda^{-1}),\\
M_{ik}=(x_i-x_k)^2+(y_i-y_k)^2,\quad i=0,\dots,N,\quad k=1,\dots,N.
\end{gathered}
$$


\paragraph*{Qualitative analysis of the relative motion for $N=1$.}

According to \eqref{eqq2}, the domains of the variables $\psi$, $\rho$ are
determined by the inequalities
$$
0\le \psi<4\pi,\quad c\le \rho.
$$

Each point of this half-strip corresponds to a pair of possible mutual
arrangements of the elliptic and point vortices \eqref{eq6}, corresponding
to $\lambda<1$ and $1<\lambda$ (see Fig.~2).

It can be shown that the {\it relation
$$
\rho=c,\quad\mbox{i.\,e. }\lambda=1,
$$
corresponds to the case when elliptic vortex becomes circular (Rankin
vortex\/)}.

The trajectories of the reduced system are determined by the contour lines
of the Hamiltonian \eqref{eq5}; therefore, the trajectories of the system
in domains $\lambda<1$ and $\lambda>1$ can be obtained with the help of
the substitution
$$
\psi\to \psi+\pi,\quad \rho\to \rho.
$$
Moreover, the Hamiltonian \eqref{eq5} is $2\pi$-periodic with respect to
$\psi$. Therefore, we will restrict ourselves to the analysis of the
contour lines of the Hamiltonian $H_{-}$ in the domain
$$
0\le \psi<2\pi,\quad c\le \rho.
$$

Normalizing the variable and the integral
\begin{equation} \label{eq7}
\rho=c\widetilde{y},\quad D=4c|1+\alpha|\widetilde D
\end{equation}
and eliminating ``insignificant'' constants in~\eqref{eq5}, we find that
the trajectories of the system are determined by the level curves of the
function
\begin{equation}
\label{eq8}
\begin{gathered}
\widetilde{H}_{-}=-\ln (1+\widetilde{y})-2\alpha\ln(\widetilde
D-\widetilde{y})-\frac{2\alpha}{1+\alpha}\frac{\sqrt{\widetilde{y}^2-1}}{\widetilde
D-\widetilde{y}}\cos\psi,\quad
1+\alpha>0,\\
\widetilde{H}_{-}=-\ln (1+\widetilde{y})-2\alpha\ln(\widetilde
D+\widetilde{y})-\frac{2\alpha}{|1+\alpha|}\frac{\sqrt{\widetilde{y}^2-1}}{\widetilde
D+\widetilde{y}}\cos\psi,\quad 1+\alpha<0.
\end{gathered}
\end{equation}

According to~\cite{cccbib4}, the equations describing the dynamics of a
Kirchhoff vortex interacting with a point vortex are valid only at a
sufficiently large distance from the Kirchhoff vortex. Without specifying
the exact domain where the results thus obtained are applicable, we give
here a trajectory of the reduced system \eqref{eq5} and mark the domain
occupied by the Kirchhoff vortex. In accordance with \eqref{eqq4},
\eqref{eq7}, the elliptic vortex in the plane $\psi$, $\widetilde{y}$
occupies the domain determined by the relationships
\begin{equation}
\label{eq9}
\begin{gathered}
\widetilde
D-\widetilde{y}\le\frac{2}{1+\alpha}(\widetilde{y}-\sqrt{\widetilde{y}^2-1}\cos\psi),\quad
1+\alpha>0,\\
\widetilde
D+\widetilde{y}\le\frac{2}{|1+\alpha|}(\widetilde{y}-\sqrt{\widetilde{y}^2-1}\cos\psi),\quad
1+\alpha<0.
\end{gathered}
\end{equation}
The domain occupied by the Kirchhoff vortex is shaded in the figures
below.

{\small
\paragraph*{Stability of the circular vortex.}

Let us make the canonical change of variables
\begin{equation}
\label{ccceqstar} \widetilde y=1+\frac{u^2+v^2}2,\quad
\psi=\arctan\frac{u}{v}.
\end{equation}
Now the Hamiltonian \eqref{eq8} near $\widetilde y=1$ can be written as
$$
\begin{gathered}
\widetilde H={\rm const}-\frac{2\sqrt2\alpha}{(1+\alpha)(\widetilde
D-1)}v-\frac12\frac{\widetilde
D-1-4\alpha}{\widetilde D-1}(u^2+v^2)+\dots,\quad1+\alpha>0;\\
\widetilde H={\rm const}+\frac{2\sqrt2\alpha}{(1+\alpha)(\widetilde
D+1)}v-\frac12\frac{\widetilde D+1+4\alpha}{\widetilde
D+1}(u^2+v^2)+\dots,\quad1+\alpha<0.
\end{gathered}
$$
Thus, after the identification of \eqref{ccceqstar}, we obtain that the
origin of coordinates $u=v=0$ is not a fixed point, i.\,e., in the
presence of point vortex, the circular vortex is locally unstable with
respect to elliptic deformations.
}

Let us qualitatively describe the structure of the phase portrait on the
$\psi$, $\widetilde{y}$ plane for different values of the parameters
$\alpha$, $\widetilde D$. As follows from \eqref{eqq4} and \eqref{eq8},
the phase space of the reduced system and the respective phase portraits
change depending on the sign of $(1+\alpha)$. Let us consider each case
separately.

\subparagraph*{${\bf 1+\alpha>0}$ (Fig.~3).} In this case, the
phase space is the rectangle on the plane $\psi$, $\widetilde{y}$:
\begin{equation}
\label{eq1103-1} 0\le \psi<2\pi,\quad 1\le \widetilde y<\widetilde D.
\end{equation}
The opposite sides of this rectangle $\psi=0$ and $\psi=2\pi$ are
identified with one another. The segment $\widetilde y=1$ corresponds to
the case of a circular Kirchhoff vortex, and $\widetilde y=\widetilde D$
is the case when the point vortex is in the center of the elliptic vortex.
In both cases the definition of the angle $\psi$ makes no sense (i.\,e.,
this is a singularity similar to the origin for polar coordinates). In
this case, the phase space can be identified with a two-dimensional
sphere.

{\sc Remark 1} The identification for the sphere
$\xi_1^2+\xi_2^2+\xi_3^2=1$ embedded into~${\mathbb{R}}^3$ can be
explicitly specified, for example, in the following manner
\begin{equation}
\label{zv12} \xi_3=\frac{2\widetilde y-\widetilde D-1}{\widetilde
D-1},\quad \xi_1=\sqrt{1-\xi_3^2}\cos \psi,\quad\xi_2=\sqrt{1-\xi_3^2}\sin
\psi.
\end{equation}


Fig.~3 gives a characteristic appearance of phase portraits
corresponding to the three different parameter domains. The appearance of
the portrait is completely determined by the critical points of the
Hamilton function \eqref{eq8}, which, as follows from \eqref{eq8}, lie on
the straight lines $\psi=0$, $\psi=\pi$. In this case the point vortex
lies on the principal axes of the elliptic Kirchhoff vortex.

Each straight line $\psi=0$ and $\psi=\pi$ (in the nondegenerate case)
contains either 0 or 2 critical points of the function \eqref{eq8}.

The domain \eqref{eq1103-1} can contain either 2 or 0 (nondegenerate)
critical points of the function \eqref{eq8} that do not lie on the
segments $\widetilde y=1$, $\widetilde y=\widetilde D$. Both critical
points lie either on the straight line $x=0$ or on the straight line
$x=\pi$. The critical point nearest to the segment $\widetilde y=1$ always
corresponds to a stable fixed point of the reduced system.

A criterion for the global stability of the Kirchhoff vortex at
$1+\alpha>0$ {\rm can be formulated within the framework of the model
considered:

if the Hamilton function has critical points in the domain
\eqref{eq1103-1}, the perturbations of the Rankin vortex remain
constrained.

Indeed, in this case, as can be seen from Fig.~3, there
always exists an invariant curve that limits the extent of the Kirchhoff
vortex deformation and inhibits its merging with the point vortex.}


\subparagraph*{${\bf 1+\alpha<0}$ (Fig.~4).} In this case, the domain
of motion in the~$\psi$, $\widetilde{y}$-plane is noncompact:
$$
0\le \psi<2\pi,\quad\widetilde y\ge\max\left(1,-\widetilde
D\right)=\widetilde y_0.
$$
Carrying out the (canonical) change of coordinates
$$
\widetilde y=\widetilde y_0+\frac{u^2+v^2}2,\quad \psi=\arctan\frac uv,
$$
we find that the phase space is the $(u,v)$-plane and the straight line
$\widetilde y=\widetilde y_0$ corresponds to the origin of coordinates.
The structure of the phase portrait (the shape of the domain occupied by
the Kirchhoff vortex) depends on the sign of $\widetilde D$.

When $\widetilde D>0$, there is a stable periodic solution on the straight
line $\psi=\pi$ (Fig.~4{\it a}). For $\widetilde D<0$, an
unstable periodic solution exists at $\psi=0$ (Fig.~4{\it b}).

\subparagraph*{$1+\alpha=0$ (the case of a vortex pair).} The
transformation~\eqref{eq7} cannot be made in this case, therefore we carry
out the normalization
\begin{equation}
\label{eq10} \rho=c\widetilde{y},\quad D=4c\widetilde
D=\Gamma_0M_{10}={\rm const}.
\end{equation}
Thus, in this case, the distance between vortices remains constant
~$(M_{10}={\rm const})$, and changes take place only in the mutual
arrangement of the vortices. The variables $\psi$, $\widetilde y$ are
determined only in the half-strip
$$
0\le \psi\le2\pi,\quad1<\widetilde y.
$$
In this case the trajectories are determined by the contour lines of the
Hamiltonian $H_{-}$ in \eqref{eq5}, which, after the elimination of
constant values can be written as
\begin{equation}
\label{eq11} \widetilde{H}_{-}=-\ln (1+\widetilde{y})+\frac{2}{\widetilde
D}\sqrt{\widetilde{y}^2-1}\cos\psi,
\end{equation}
The domain occupied by the elliptic vortex on the domain of the variables
is determined by the inequality
\begin{equation}
\label{eq12} \widetilde D\ge
2(\widetilde{y}-\sqrt{\widetilde{y}^2-1}\cos\psi).
\end{equation}
Whence it follows that for $\widetilde D<0$ the entire plane
$(\psi,\widetilde y)$ is occupied by the elliptic vortex; therefore, we
assume that $\widetilde D>0$.


For $\widetilde D<\widetilde D_{*}=\sqrt{22+10\sqrt5}$, the phase portrait
contains no periodic solutions (Fig.~5{\it a}).

When $\widetilde D>\widetilde D_{*}$, two periodic solutions appear in the
phase portrait on the axis $\psi=0$; one of these solutions is stable,
while the other is unstable (Fig.~5{\it b,c})

\section{Interaction between two Kirchhoff vortices }

\subparagraph*{Reduction to a system with two degrees of freedom.} The
dynamics of two Kirchhoff vortices can be described by the
Hamiltonian~\eqref{in3}, which can be represented in the form
\begin{equation}
\label{eqt1}
\begin{gathered}
H=H_1+H_2+H_3,\\
H_1=-\frac{\Gamma_1^2}{8\pi}\ln
\frac{(1+\lambda_1)^2}{4\lambda_1}-\frac{\Gamma_2^2}{8\pi}\ln
\frac{(1+\lambda_2)^2}{4\lambda_2},\\
H_2=-\frac{\Gamma_1\Gamma_2}{4\pi}\ln M,\\
H_3=-\frac{\Gamma_1\Gamma_2}{16\pi^2M}\left(\frac{S_1(1-\lambda_1^2)}{\lambda_1}
\cos{2(\theta-\varphi_1)}-\frac{S_2(1-\lambda_2^2)}{\lambda_2}
\cos{2(\theta-\varphi_2)}\right),\\
M=(x_1{-}x_2)^2{+}(y_1{-}y_2)^2,\quad
\theta=\theta_{12}=\pi{+}\theta_{21}=\arctan{\frac{y_2{-}y_1}{x_2{-}x_1}}.
\end{gathered}
\end{equation}

The particular solution of the system \eqref{eqt1}, at which
$\gamma_1=\gamma_2$ and the vortices are centrally symmetrical with
respect to one another is suggested and studied by Melander {\em et al.} \cite{cccbib11}. In
that paper, the conditions for merging of two vortex patches are given.
These conditions, within the framework of the moment model, were found to
be equivalent to the collapse of two Kirchhoff vortices, during which
their centers coincide after a finite time.

Let us consider the system of relative variables
\begin{equation}
\label{eq2}
\begin{gathered}
\psi_i=2(\theta-\varphi_i),\quad
\rho_i=\frac{c_i}{2}\left(\lambda_i+\frac1{\lambda_i}\right),\quad
c_i=\frac{\Gamma_iS_i}{8\pi},\quad i=1,2.\\
z=\frac14M.
\end{gathered}
\end{equation}

These variables commute with the integrals~\eqref{in4}; they are closed
with respect to the Poisson brackets~\eqref{in1}, and their commutation
relationships have the form
\begin{equation}
\label{eqt3} \{\psi_i,\rho_i\}=\delta_{ij},\;
\{\psi_i,z\}=-(\Gamma_1^{-1}+\Gamma_2^{-1}),\;\{\rho_i,z\}=0,\; i,j=1,2.
\end{equation}

The Poisson structure~\eqref{eqt3} has a linear Kazimir function (integral
of motion)
\begin{equation}
\label{eqt4} D=z+(\Gamma_1^{-1}+\Gamma_2^{-1})(\rho_1+\rho_2).
\end{equation}

As follows from~\eqref{eqt3},~\eqref{eqt4}, in the case of the vortex
pair~$(\Gamma_1+\Gamma_2\!=\!0)$, the distances between the vortex centers
remain constant.

Eliminating $z$ using the integral~\eqref{eqt4} and expressing the
Hamiltonian~\eqref{eqt1} in terms of the variables $\psi_1$, $\rho_1$, we
obtain the Hamiltonian of the reduced canonical system with two degrees of
freedom
\begin{equation}
\begin{gathered}
H=-\frac{\Gamma_1^2}{8\pi}\ln (c_1+\rho_1)-\frac{\Gamma_2^2}{8\pi}\ln
(c_2+\rho_2)\\
-\frac{\Gamma_1\Gamma_2}{4\pi}\ln
\left(D-(\Gamma_1^{-1}+\Gamma_2^{-1})(\rho_1+\rho_2)\right) \label{qwerr}\\
-\frac{\Gamma_1\Gamma_2}{4\pi(D-(\Gamma_1^{-1}+\Gamma_2^{-1})(\rho_1+\rho_2))}\left(\pm
\frac{\sqrt{\rho_1^2-c_1^2}}{\Gamma_1}\cos{\psi_1}\mp
\frac{\sqrt{\rho_2^2-c_2^2}}{\Gamma_2}\cos{\psi_2}\right)\!\!,
\end{gathered}
\end{equation}
where the upper sign corresponds to the condition $\lambda_i>1$, and the
lower sign to the condition $0<\lambda_i\le 1$ (see.~\eqref{eq6}).

\paragraph*{Absolute motion.}

Let us find the equations determining the positions $(x_1,y_1)$ and
$(x_2,y_2)$ and orientation $\varphi_1$, $\varphi_2$ of two Kirchhoff
vortices in stationary space if their relative arrangement determined by
the system \eqref{qwerr} is assumed to be known: $\rho_i=\rho_i(t)$,
$\psi_i=\psi(t)$, $i=1,2$. Direct calculations show that the slopes of the
principal  (major) semiaxes of the ellipse are determined by the
quadratures
\begin{equation}
\label{ccceqa1}
\dot\varphi_i=\frac{\Gamma_ic_i}{S_i}\frac{\rho_i+\sqrt{\rho_i^2-c_i^2}}{\left(
\rho_i+c_i+\sqrt{\rho_i^2-c_i^2}\right)^2}-\frac{\Gamma_1\Gamma_2}{2\pi
\Gamma_i
M}\frac{\left(\rho_i+\sqrt{\rho_i^2-c_i^2}\right)\rho_i\cos\psi_i}
{\rho_i^2-c_i^2+\rho_i\sqrt{\rho_i^2-c_i^2}}.
\end{equation}
If $\Gamma_1+\Gamma_2\ne0$, the positions of the vortex centers can be
found from the system of linear equations with coefficients explicitly
depending on time:
\begin{equation}
\label{ccceqa2}
\begin{gathered}
x_1=\frac{Q+\Gamma_2\Delta x}{\Gamma_1+\Gamma_2},\quad
x_2=\frac{Q-\Gamma_1\Delta x}{\Gamma_1+\Gamma_2},\quad
y_1=\frac{P+\Gamma_2\Delta y}{\Gamma_1+\Gamma_2},\quad
y_2=\frac{P-\Gamma_2\Delta y}{\Gamma_1+\Gamma_2};\\
\Delta\dot x=-\frac{\Gamma_1+\Gamma_2}{\pi^2M^2}\left(\frac{\pi M}2\Delta
y+\frac{S_1\sqrt{\rho_1^2-c_1^2}}{4c_1}(\Delta x\sin\psi_1+\Delta
y\cos\psi_1)\right.\\
\left.\qquad\quad+\frac{S_2\sqrt{\rho_2^2-c_2^2}}{4c_2}(\Delta
x\sin\psi_2+\Delta
y\cos\psi_2)\right);\\
\Delta\dot y=\frac{\Gamma_1+\Gamma_2}{\pi^2M^2}\left(\frac{\pi M}2\Delta
x+\frac{S_1\sqrt{\rho_1^2-c_1^2}}{4c_1}(\Delta x\cos\psi_1-\Delta
y\sin\psi_1)\right.\\
\left.\qquad\quad+\frac{S_2\sqrt{\rho_2^2-c_2^2}}{4c_2}(\Delta
x\cos\psi_2-\Delta y\sin\psi_2)\right)
\end{gathered}
\end{equation}
where $\Delta x=x_1-x_2$, $\Delta y=y_1-y_2$, and $Q$, $P$ are integrals
\eqref{in4}.

\paragraph*{Integrable case $\Gamma_1+\Gamma_2=0$.}
In the case of the vortex pair $\Gamma_1=-\Gamma_2$, the
Hamiltonian~\eqref{qwerr} can be split into two independent Hamiltonians
$$
H=H_1(\psi_1,\rho_1)+H_2(\psi_2,\rho_2),
$$
and the system can be integrated by the method of separation of variables.
Thus, we obtain a new nontrivial integrable case for the vortex dynamics.
This case of integrability was first mentioned in our review
\cite{cccbib1}. As follows from equations \eqref{ccceqa2}, in this case
$$
x_1-x_2={\rm const},\quad y_1-y_2={\rm const},
$$
and hence
$$\theta=\arctan\frac{y_2-y_1}{x_2-x_1}={\rm const}.$$
Thus, the
length and orientation of the segment connecting the centers of vortices
remain unchanged during the motion (see Fig.~1).

\paragraph*{Poincar\'{e} sections and integrability.}
The system \eqref{qwerr} is not integrable in the general case. This is
demonstrated by the chaotic trajectories constructed at
$\Gamma_1=\Gamma_2$ with the use of a Poincar\'{e} section shown in Fig.~6.
It should be mentioned that there are few returning
trajectories in the case of interaction between Kirchhoff vortices, which
hampers the numerical analysis. Fig.~7 gives numerically
plotted separatrices of a hyperbolic fixed point of the Poincar\'{e}
mapping. Their transversal intersection (established by computer analysis)
is an indication that the problem of Kirchhoff vortex motion is
nonintegrable. This fact was mentioned as a supposition in \cite{cccbib4}.
No analytical proof of the nonintegrability of two Kirchhoff vortices
within the framework of the second-order moment model has been obtained
yet.


\section{Acknowledgements} This work was supported by RFBR (04-05-64367 and
05-01-01058), CRDF (RU-M1-2583-MO-04), INTAS (04-80-7297) and the program
``State Support for Leading Scientific Schools'' (136.2003.1).









\newpage

\begin{figure}
\begin{center}
\includegraphics{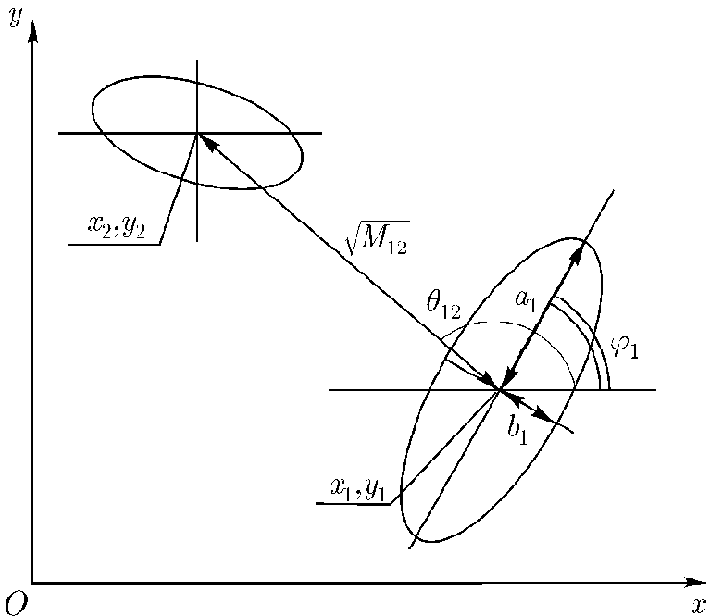}
\caption{Elliptic vortex patches}
\end{center}
\label{ell}
\end{figure}

\begin{figure}
\begin{center}
\includegraphics{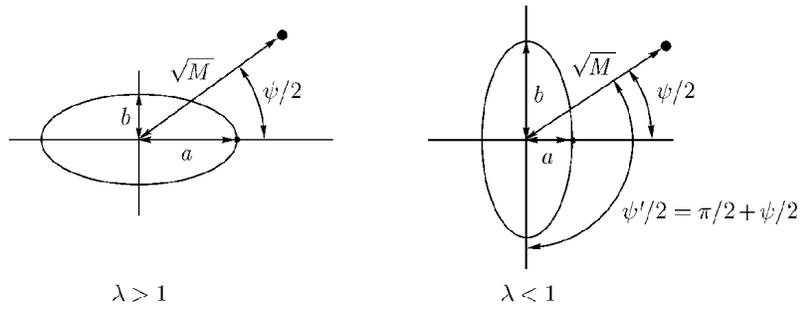}
\caption{Mutual arrangement of the elliptic and point vortices with respect to a chosen point on the contour of the elliptic vortex in the case where $\lambda<1$ and
$\lambda>1$.}
\end{center}
\label{2new}
\end{figure}

\begin{figure}
\begin{center}
\includegraphics{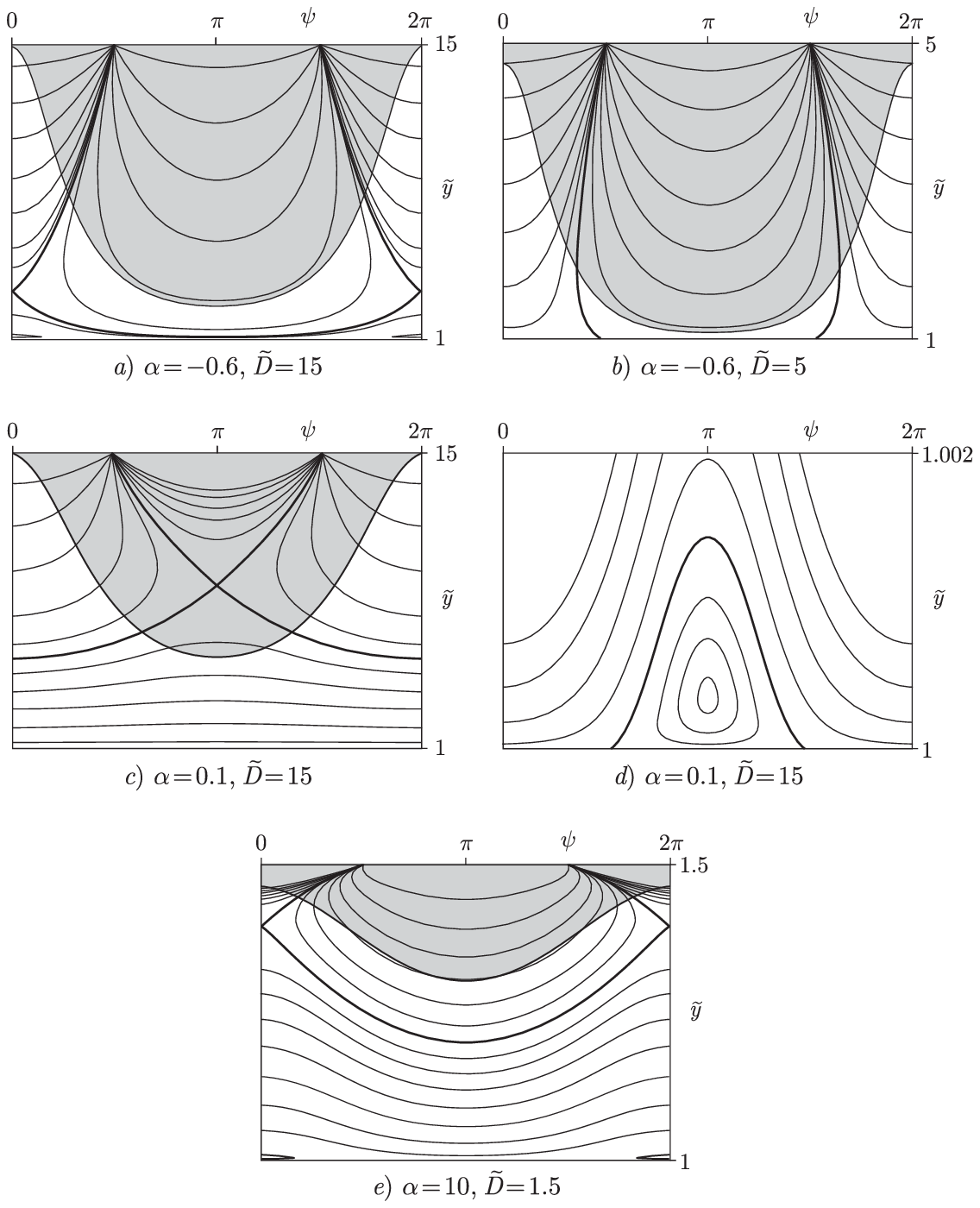}
\caption{Phase portraits for $1+\alpha>0$ (the domain occupied by the
Kirchhoff vortex is colored gray). Figure $3d$ is an enlarged part of
figure $3c$ near the lower segment.}
\end{center}
\end{figure}

\begin{figure}
\begin{center}
\includegraphics{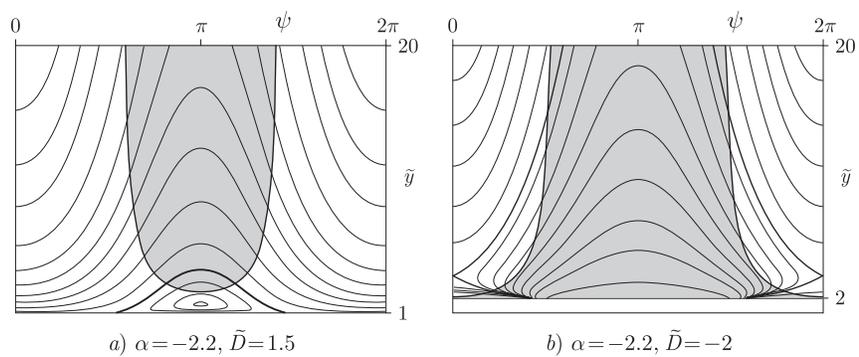}
\end{center}
\caption{Phase portraits for $1+\alpha<0$ (the domain occupied by the
Kirchhoff vortex is colored gray).}
\end{figure}

\begin{figure}
\begin{center}
\includegraphics{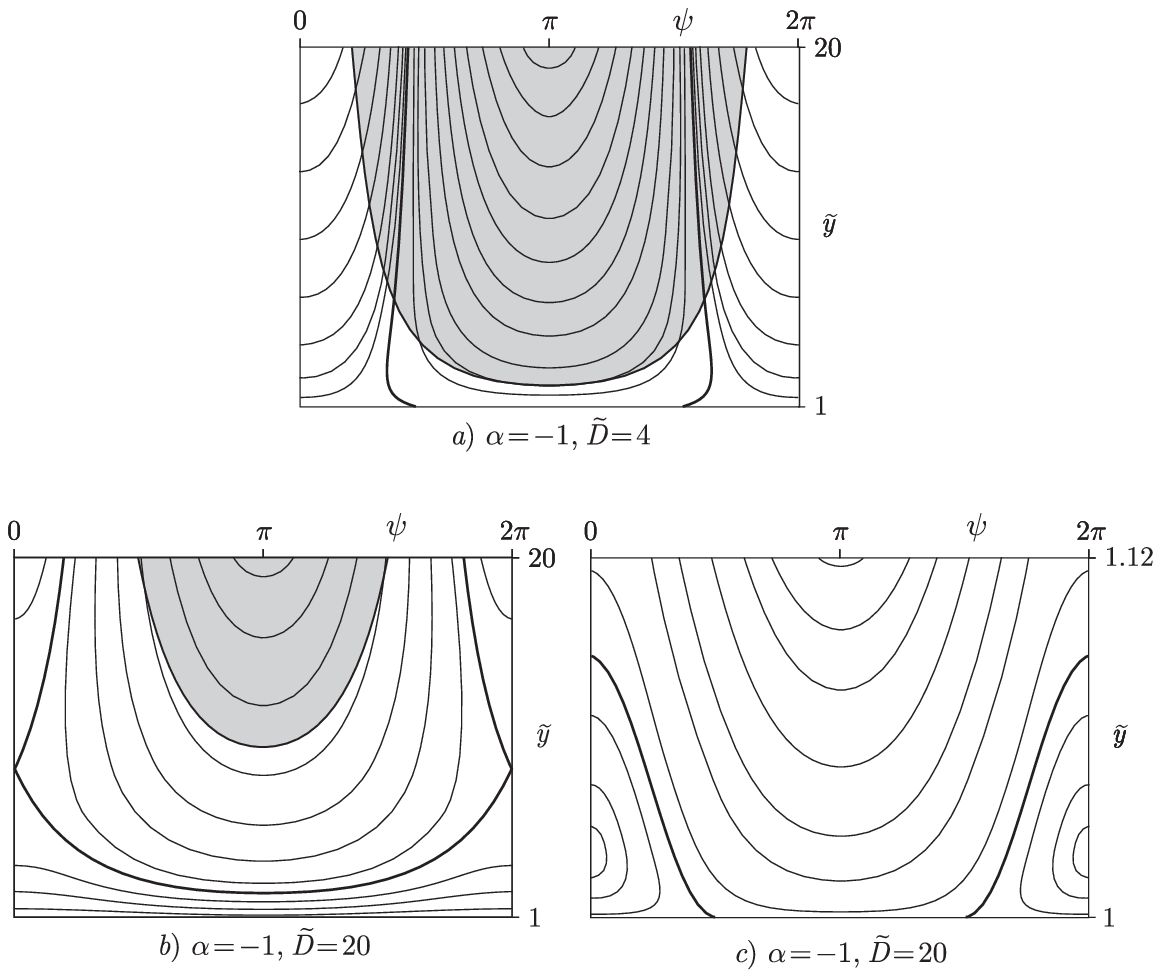}
\caption{Phase portrait for $1+\alpha=0$ (the domain occupied by the
Kirchhoff vortex is colored gray). Figure $5c$ is an enlarged part of
figure $5b$ near the lower segment.}
\end{center}
\end{figure}

\begin{figure}
\begin{center}
\includegraphics{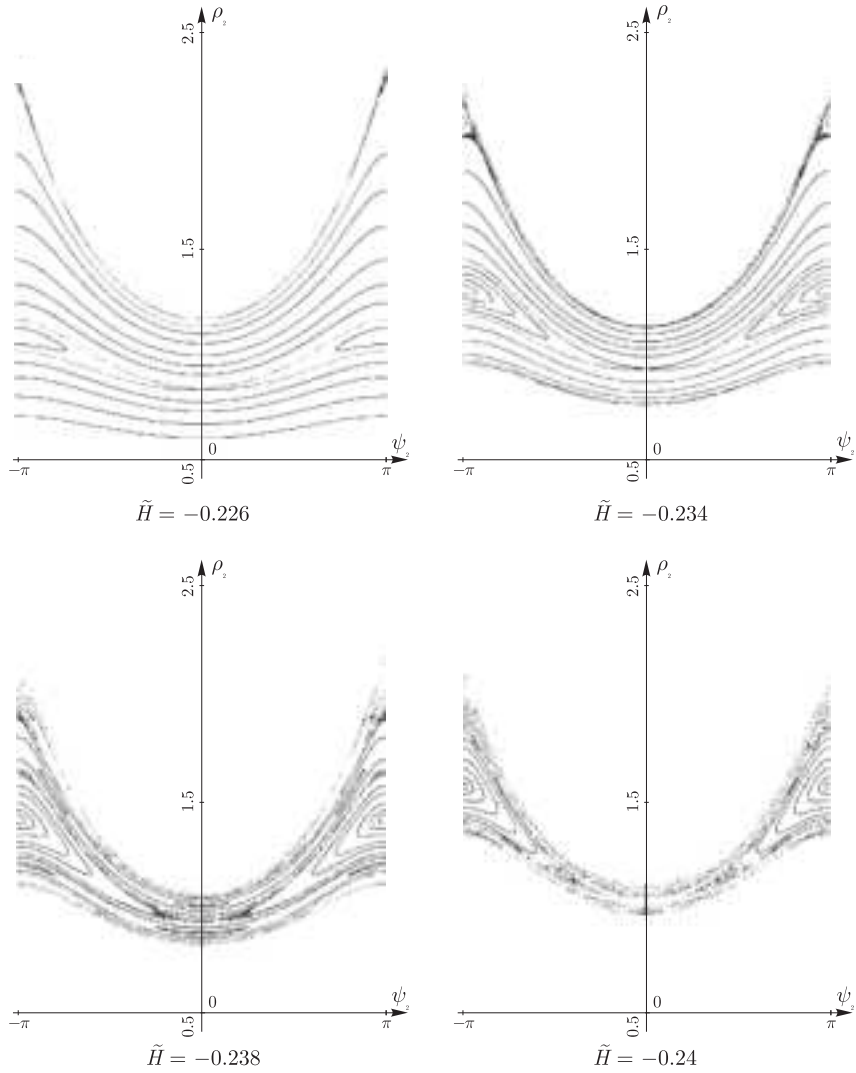}
\caption{Poincar\'{e} mapping for the reduced system \eqref{qwerr} in
the problem of two elliptic vortices. The section by the plane $\psi_1=0$
was chosen for the following parameter values: $\gamma_1=\gamma_2=1$,
$S_1=S_2=0.3$ and $D=22$.}
\end{center}
\end{figure}

\begin{figure}
\begin{center}
\includegraphics{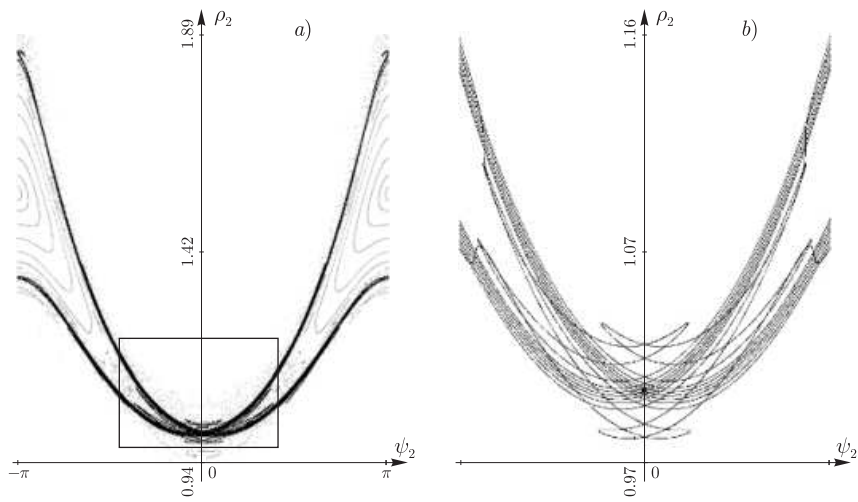}
\caption{Separatrices for the last phase portrait shown in Fig.~6
$(\widetilde H=-0.24)$. Fig.~7$b$ is the enlarged central
part of figure 7$a$.}
\end{center}
\end{figure}

\end{document}